\documentclass{article}
\usepackage{amsfonts}
\usepackage{amsmath}

\input{tcilatex}

\begin{document}

\title{Simultaneous eigenstates of the number-difference operator and a
bilinear interaction Hamiltonian derived by solving a complex differential
equation\thanks{%
Work supported by The Presidential Foundation of the Chinese Academy of
Science}}
\author{$^{1}$Hong-yi Fan and $^{2}$Wei-bo Gao \\
$^{1}$Department of Material Science and Engineering,\\
University of Science and Technology of China, Hefei, Anhui 230026, China\\
$^{2}$Department of Modern Physics, University of Science and Technology\\
of China, Hefei, Anhui 230026, China}
\maketitle

\begin{abstract}
As a continuum work of Bhaumik et al who derived the common eigenvector of
the number-difference operator $Q\equiv \left( a^{\dagger }a-b^{\dagger
}b\right) $ and pair-annihilation operator $ab$ (J. Phys. A9 (1976) 1507) we
search for the simultaneous eigenvector of $Q$ and $\left( ab-a^{\dagger
}b^{\dagger }\right) $ by setting up a complex differential equation in the
bipartite entangled state representation. The differential equation is then
solved in terms of the two-variable Hermite polynomials and the formal
hypergeometric functions. The work is also an addendum to Mod. Phys. Lett. A
9 (1994) 1291 by Fan and Klauder, in which the common eigenkets of $Q$ and
pair creators $a^{\dagger }b^{\dagger }$ are discussed.
\end{abstract}

\section{Introduction}

Coherent states [1-2] are widely used in many aspects of quantum physics.
The bosonic coherent state, which describes the quantum state of laser,
obeys coordinate-momentum minimum uncertainty relation and possesses
non-orthonormal and over-complete properties. Generalized coherent states
has been constructed and applied by theoreticians since 1970s, among which
the coherent state $\left\vert q,\alpha \right\rangle $ for charged bosons
[3] is a remarkable one, when one introduces "charge" by defining two types
of quanta possessing "charge" +1 and -1 with corresponding annihilation
operators\ $a$ and $b,$ so the operator $Q=a^{\dagger }a-b^{\dagger }b$ is
endowed with charge operator, $\left\vert q,\alpha \right\rangle $ is
constructed based on $\left[ Q,ab\right] =0,$ $\left[ a,a^{\dagger }\right] =%
\left[ b,b^{\dagger }\right] =1.$ In the Fock space the charged coherent
state is 
\begin{eqnarray*}
\left\vert q,\alpha \right\rangle  &=&C_{q}\sum_{n=0}^{\infty }\frac{\alpha
^{q}}{\sqrt{(n+q)!n!}}\left\vert n+q,n\right\rangle , \\
Q\left\vert q,\alpha \right\rangle  &=&q\left\vert q,\alpha \right\rangle ,%
\text{\ \ }ab\left\vert q,\alpha \right\rangle =\alpha \left\vert q,\alpha
\right\rangle 
\end{eqnarray*}%
where $C_{q}$ is the normalization constant. In quantum optice theory $%
\left\vert q,\alpha \right\rangle $ was named by Agarwal as pair-coherent
state\ in [4], $Q$ is the two-mode photon number-difference operator. By
observing $\left[ Q,a^{\dagger }b^{\dagger }\right] =0,$ in Ref. [5] Fan and
Klauder also constructed the common eigenvector of $Q$ and $a^{\dagger
}b^{\dagger }$ with use of the Dirac's $\delta $-function in countor
integral form proposed by Heitler [6]-[7]. One then naturally to ask waht is
the simultaneous eigenstates, denoted as $\left\vert q,k\right\rangle ,$ of $%
Q$ and $\left( ab-a^{\dagger }b^{\dagger }\right) ?$ This question is full
of physical meaning in quantum optics, because most nonlinear interactions
in tha parametric approximation reduce to a bilinear form 
\begin{equation}
H_{I}=i\hbar \kappa \left( ab-a^{\dagger }b^{\dagger }\right) ,  \label{1}
\end{equation}%
since 
\begin{equation}
\left[ \left( ab-a^{\dagger }b^{\dagger }\right) ,Q\right] =0,  \label{2}
\end{equation}%
where $\kappa $ is related to the susceptibility in the parametric process. $%
H_{I}$ is responsible for producing two-mode squeezed states [8-11], which
is not only useful for optical communication and weak signal detection, but
also embodies quantum entanglement, i.e. the Einstein-Podolsky-Rosen (EPR)
correlations [12] for quadrature phase amplitude are intrinsic to two-mode
squeezed light, the idler-mode and the signal-mode generated from a
parametric amplifier are entangled each other in a frequency domain. Solving
this question is quite difficult. We recall Dirac's guidence [13]:
\textquotedblleft When one has a particular problem to work out in quantum
mechanics, one can minimize the labour by using a represenation in which the
representatives of the more important abstract quantities occurring in that
problem are as simple as possible\textquotedblright . At first glance, we
thought that the charged coherent state representation $\left\vert q,\alpha
\right\rangle $ was a good candidates for tackling the problem as simple as
possible, but after some tries we found that it was not. Eventually we find
that the entangled state representation [14-15] is of assistence for
searching for the desired common eigenvector $\left\vert q,k\right\rangle $
of $Q$ and $\left( ab-a^{\dagger }b^{\dagger }\right) $, 
\begin{equation}
Q\left\vert q,k\right\rangle =q\left\vert q,k\right\rangle ,\text{ }
\label{3}
\end{equation}%
\begin{equation}
\left( ab-a^{\dagger }b^{\dagger }\right) \left\vert q,k\right\rangle
=\left( q-k-1\right) \left\vert q,k\right\rangle .  \label{4}
\end{equation}%
because it possesses well-behaved features. Thus we shall briefly review the
properties of entangled state $\left\vert \xi \right\rangle $ in Sec. 2.
Then in Sec. 3 we shall make full use of $\left\vert \xi \right\rangle $ to
set up a complex differential equation for the overlap $\left\langle \xi
\right\vert \left. q,k\right\rangle .$ In Sec. 4 we employ the two-variable
Hermite polynomials and the hypergeometric function to solve the
differential equation. The Gauss' contiguous relation of hypergeometric
function is essential for us to derive the common eigenvector of $Q$ and $%
\left( ab-a^{\dagger }b^{\dagger }\right) $.

\section{Brief review of the bipartite entangled state representations}

In [14] and [15] we have constructed the bipartite entangled state%
\begin{equation}
\exp \left[ -\frac{|\xi |^{2}}{2}+\xi a^{\dagger }+\xi ^{\ast }b^{\dagger
}-a^{\dagger }b^{\dagger }\right] \left\vert 0\right\rangle \equiv
\left\vert \xi \right\rangle ,\text{ }\xi =|\xi |e^{i\varphi },  \label{5}
\end{equation}%
$\left\vert \xi \right\rangle $ satisfy the eigenvator equations%
\begin{equation}
\left( a+b^{\dagger }\right) \left\vert \xi \right\rangle =\xi \left\vert
\xi \right\rangle ,\ \left( a^{\dagger }+b\right) \left\vert \xi
\right\rangle =\xi ^{\ast }\left\vert \xi \right\rangle .  \label{6}
\end{equation}%
or 
\begin{equation}
\frac{1}{2}\left( X_{1}+X_{2}\right) \left\vert \xi \right\rangle =\frac{1}{%
\sqrt{2}}\xi _{1}\left\vert \xi \right\rangle ,\text{ \ }\left(
P_{1}-P_{2}\right) \left\vert \xi \right\rangle =\sqrt{2}\xi _{2}\left\vert
\xi \right\rangle ,  \label{7}
\end{equation}%
i.e. $\left\vert \xi \right\rangle $ is just the simultaneous eigenvector of
two-particles' center-of-mss posititon $X_{1}-X_{2}$ and the relative
momentum $P_{1}-P_{2}$ in Fock space, we name it the EPR eigenstate since
EPR were the first who used the commutative property $\left[
X_{1}-X_{2},P_{1}+P_{2}\right] =0$ to challege the incompleteness of quantum
mechanics [11]. According to Dirac's representation theory $[13]$:
\textquotedblleft To set up a representation in a general way, we take a
complete set of bra vectors, i.e. a set such that any bra can be expressed
linearly in terms of them,\textquotedblright\ we use the normal ordering
form of the two-mode vacuum state projector 
\begin{equation}
\left\vert 00\right\rangle \left\langle 00\right\vert =:e^{-a^{\dagger
}a-b^{\dagger }b}:,  \label{8}
\end{equation}%
and the technique of integration within an ordered product (IWOP) of
operators [16]-[17] , we can prove 
\begin{equation}
\int \frac{d^{2}\xi }{\pi }\left\vert \xi \right\rangle \left\langle \xi
\right\vert =1.  \label{9}
\end{equation}

\section{Complex differential equation for the new state $\left\vert
q,k\right\rangle $}

By noticing (5) we see%
\begin{eqnarray}
a\left\vert \xi \right\rangle &=&\left( \xi -b^{\dagger }\right) \left\vert
\xi \right\rangle ,\text{\ \ }b\left\vert \xi \right\rangle =\left( \xi
^{\ast }-a^{\dagger }\right) \left\vert \xi \right\rangle ,  \label{10} \\
\left\langle \xi \right\vert a^{\dagger } &=&\left\langle \xi \right\vert
\left( \xi ^{\ast }-b\right) ,\text{\ \ }\left\langle \xi \right\vert
b^{\dagger }=\left\langle \xi \right\vert \left( \xi -a\right) ,  \notag
\end{eqnarray}%
so%
\begin{equation}
\left\langle \xi \right\vert \left( ab-a^{\dagger }b^{\dagger }\right)
=\left\langle \xi \right\vert \left[ ab-\left( \xi ^{\ast }-b\right)
b^{\dagger }\right] =\left\langle \xi \right\vert \left[ b\xi -\left( \xi
-a\right) \xi ^{\ast }+1\right] .  \label{11}
\end{equation}%
Then using 
\begin{equation}
a^{\dagger }\left\vert \xi \right\rangle =\left( \frac{\partial }{\partial
\xi }+\frac{\xi ^{\ast }}{2}\right) \left\vert \xi \right\rangle ,\text{\ \ }%
b^{\dagger }\left\vert \xi \right\rangle =\left( \frac{\partial }{\partial
\xi ^{\ast }}+\frac{\xi }{2}\right) \left\vert \xi \right\rangle ,
\label{12}
\end{equation}%
we re-write (11) as 
\begin{equation}
\left\langle \xi \right\vert \left( ab-a^{\dagger }b^{\dagger }\right)
=\left\langle \xi \right\vert \left[ \left( \frac{\overleftarrow{\partial }}{%
\partial \xi }+\frac{\xi ^{\ast }}{2}\right) \xi +\left( \frac{%
\overleftarrow{\partial }}{\partial \xi ^{\ast }}+\frac{\xi }{2}\right) \xi
^{\ast }-|\xi |^{2}+1\right] .  \label{13}
\end{equation}%
Operating (13) on the state $\left\vert q,k\right\rangle $, and using (4) we
obtain a complex differential equation 
\begin{eqnarray}
&&\left\langle \xi \right\vert \left( ab-a^{\dagger }b^{\dagger }\right)
\left\vert q,k\right\rangle  \label{14} \\
&=&\left[ \xi \frac{\partial }{\partial \xi }+\xi ^{\ast }\frac{\partial }{%
\partial \xi ^{\ast }}+1\right] \left\langle \xi \right\vert \left.
q,k\right\rangle =\left( q-k-1\right) \left\langle \xi \right\vert \left.
q,k\right\rangle  \notag
\end{eqnarray}%
On the other hand, from (5) we have 
\begin{eqnarray}
\left( a^{\dagger }a-b^{\dagger }b\right) \left\vert \xi \right\rangle
&=&\left( \xi a^{\dagger }-b^{\dagger }\xi ^{\ast }\right) \left\vert \xi
\right\rangle  \label{15} \\
&=&|\xi |\left( e^{-i\varphi }a^{\dagger }-e^{i\varphi }b^{\dagger }\right)
\exp \left[ -\frac{|\xi |^{2}}{2}+|\xi |\left( e^{-i\varphi }a^{\dagger
}+e^{i\varphi }b^{\dagger }\right) -a^{\dagger }b^{\dagger }\right]
\left\vert 00\right\rangle  \notag \\
&=&-i\frac{\partial }{\partial \varphi }\left\vert \xi \right\rangle ,\text{%
\ \ }  \notag
\end{eqnarray}%
which together with (3) leads to another differential equation%
\begin{equation}
\left\langle \xi \right\vert \left( a^{\dagger }a-b^{\dagger }b\right)
\left\vert q,k\right\rangle =i\frac{\partial }{\partial \varphi }%
\left\langle \xi \right\vert \left. q,k\right\rangle =q\left\vert
q,k\right\rangle .  \label{16}
\end{equation}%
In the next section we want to solve (14) and (16).

\section{Solution to Eq. (16)}

After many tries we find the solution of (14) and (16) is%
\begin{equation}
\left\langle \xi \right\vert \left. q,k\right\rangle =\mathfrak{C}%
(k)e^{-|\xi |^{2}/2}A,  \label{17}
\end{equation}%
where $\mathfrak{C}(k)$ is the normalization constant, which keeps
undisturbed in the following calculations, so we neglect it in the
derivation process, and  
\begin{equation}
A\equiv \sum_{n=0}^{\infty }\frac{1}{n!}H_{n+q,n}\left( \xi ^{\ast },\xi
\right) \text{ }_{2}F_{1}(-n,\frac{k}{2}+1;q+1;2)  \label{18}
\end{equation}%
$_{2}F_{1}$ is the hypergeometric function defined as%
\begin{equation}
_{2}F_{1}(\alpha ,\beta ;\gamma ;\varepsilon )=\sum_{n=0}^{\infty }\frac{%
\left( \alpha \right) _{n}\left( \beta \right) _{n}}{\left( \gamma \right)
_{n}}\frac{z^{n}}{n!}=\frac{\Gamma (\gamma )}{\Gamma (\alpha )\Gamma (\beta )%
}\sum_{n=0}^{\infty }\frac{\Gamma (n+\alpha )\Gamma (n+\beta )}{\Gamma
(n+\gamma )}\frac{z^{n}}{n!},  \label{19}
\end{equation}
the symbol $\left( \alpha \right) _{n}$ means 
\begin{equation}
\left( \alpha \right) _{n}=\frac{\Gamma (\alpha +n)}{\Gamma (\alpha )}%
=\alpha \left( \alpha +1\right) (\alpha +2)\cdot \cdot \cdot (\alpha +n-1),
\label{20}
\end{equation}%
and the two-variable Hermite polynomials is defined as [18] 
\begin{equation}
H_{m,n}(\xi ,\xi ^{\ast })=\sum_{l=0}^{\min (m,n)}\frac{m!n!}{l!(m-l)!(n-l)!}%
(-1)^{l}\xi ^{m-l}\xi ^{\ast n-l}=H_{n,m}(\xi ^{\ast },\xi ),  \label{21}
\end{equation}%
whose generating function is%
\begin{equation}
\sum_{m,n=0}^{\infty }\frac{t^{m}t^{\prime n}}{m!n!}H_{m,n}(\xi ,\xi ^{\ast
})=\exp \left( -tt^{\prime }+t\xi +t^{\prime }\xi ^{\ast }\right) .
\label{22}
\end{equation}%
Note that the convergent condition for the hypergeometric function defined
in (19) is $|z|<1,$ $\gamma \neq 0,-1,-2,\cdot \cdot \cdot ,$ so $%
_{2}F_{1}(-n,\frac{k}{2}+1;q+1;2)$ is a formal power series. 

Now we prove (14): Firstly, we notice%
\begin{equation}
\xi \frac{\partial }{\partial \xi }\left\langle \xi \right\vert \left.
q,k\right\rangle =\xi \left( -\frac{\xi ^{\ast }}{2}e^{-|\xi
|^{2}/2}A+e^{-|\xi |^{2}/2}\frac{\partial }{\partial \xi }A\right) ,
\label{23}
\end{equation}%
and%
\begin{equation}
\xi ^{\ast }\frac{\partial }{\partial \xi ^{\ast }}\left\langle \xi
\right\vert \left. q,k\right\rangle =\xi ^{\ast }\left( -\frac{\xi }{2}%
e^{-|\xi |^{2}/2}A+e^{-|\xi |^{2}/2}\frac{\partial }{\partial \xi ^{\ast }}%
A\right) ,  \label{24}
\end{equation}%
it then follows%
\begin{equation}
\left( \xi \frac{\partial }{\partial \xi }+\xi ^{\ast }\frac{\partial }{%
\partial \xi ^{\ast }}\right) \left\langle \xi \right\vert \left.
q,k\right\rangle =-|\xi |^{2}\left\langle \xi \right\vert \left.
q,k\right\rangle +e^{-|\xi |^{2}/2}\left( \xi \frac{\partial }{\partial \xi }%
+\xi ^{\ast }\frac{\partial }{\partial \xi ^{\ast }}\right) A  \label{25}
\end{equation}%
so%
\begin{eqnarray}
&&\left( \xi \frac{\partial }{\partial \xi }+\xi ^{\ast }\frac{\partial }{%
\partial \xi ^{\ast }}+1\right) \left\langle \xi \right\vert \left.
q,k\right\rangle   \label{26} \\
&=&e^{-|\xi |^{2}/2}\left( \xi \frac{\partial }{\partial \xi }+\xi ^{\ast }%
\frac{\partial }{\partial \xi ^{\ast }}+1-|\xi |^{2}\right)
\sum_{n=0}^{\infty }\frac{1}{n!}H_{n+q,n}\left( \xi ^{\ast },\xi \right) 
\text{ }_{2}F_{1}(-n,\frac{k}{2}+1;q+1;2).  \notag
\end{eqnarray}%
Then we use the property%
\begin{equation}
\frac{\partial }{\partial \xi }H_{m,n}\left( \xi ^{\ast },\xi \right)
=nH_{m,n-1}\left( \xi ^{\ast },\xi \right) ,\text{ }\frac{\partial }{%
\partial \xi ^{\ast }}H_{m,n}\left( \xi ^{\ast },\xi \right)
=mH_{m-1,n}\left( \xi ^{\ast },\xi \right)   \label{27}
\end{equation}%
and 
\begin{equation}
H_{m+1,n}+nH_{m,n-1}=\xi ^{\ast }H_{m,n},\ H_{m,n+1}+mH_{m-1,n}=\xi H_{m,n},
\label{28}
\end{equation}%
which can be derived from (21) and (23), as well as%
\begin{eqnarray}
|\xi |^{2}H_{m,n} &=&\xi ^{\ast }\left( H_{m+1,n}+nH_{m,n-1}\right) 
\label{29} \\
&=&H_{m+1,n+1}+nmH_{m-1,n-1}+\left( m+n+1\right) H_{m,n},  \notag
\end{eqnarray}%
we have%
\begin{equation}
\begin{array}{c}
\left( \xi \frac{\partial }{\partial \xi }+\xi ^{\ast }\frac{\partial }{%
\partial \xi ^{\ast }}+1\right) \left\langle \xi \right\vert \left.
q,k\right\rangle  \\ 
=e^{-|\xi |^{2}/2}\sum_{n=0}^{\infty }\frac{1}{n!}[\xi nH_{n+q,n-1}+\xi
^{\ast }(q+n)H_{n+q-1,n} \\ 
-H_{n+q+1,n+1}-n(n+q)H_{n+q-1,n-1}-\left( q+2n\right) H_{n+q,n}]\text{ }%
_{2}F_{1}(-n,\frac{k}{2}+1;q+1;2) \\ 
=e^{-|\xi |^{2}/2}\sum_{n=0}^{\infty }\frac{1}{n!}\{n\left[
H_{n+q,n}+(n+q)H_{n+q-1,n-1}\right]  \\ 
+(q+n)[H_{n+q,n}+nH_{n+q-1,n-1}] \\ 
-H_{n+q+1,n+1}-n(n+q)H_{n+q-1,n-1}-\left( q+2n\right) H_{n+q,n}\}\text{ }%
_{2}F_{1}(-n,\frac{k}{2}+1;q+1;2) \\ 
=e^{-|\xi |^{2}/2}\sum_{n=0}^{\infty }\frac{1}{n!}%
[(q+n)nH_{n+q-1,n-1}-H_{n+q+1,n+1}]\text{ }_{2}F_{1}(-n,\frac{k}{2}+1;q+1;2)
\\ 
=e^{-|\xi |^{2}/2}\sum_{n=0}^{\infty }\frac{1}{n!}H_{n+q,n}[(q+n+1)\text{ }%
_{2}F_{1}(-n-1,\frac{k}{2}+1;q+1;2) \\ 
-n\text{ }_{2}F_{1}(-n+1,\frac{k}{2}+1;q+1;2)].%
\end{array}
\label{30}
\end{equation}%
Then using two Gauss' contiguous relations about the hypergeometric function
[19-20]%
\begin{equation}
_{2}F_{1}(\alpha ,\beta ;\gamma ;\varepsilon )=\ _{2}F_{1}(\beta ,\alpha
;\gamma ;\varepsilon ),  \label{31}
\end{equation}%
and%
\begin{equation}
\begin{array}{c}
\lbrack \gamma -2\beta +(\beta -\alpha )\varepsilon ]\text{ }%
_{2}F_{1}(\alpha ,\beta ;\gamma ;\varepsilon )+\beta (1-\varepsilon
)_{2}F_{1}(\alpha ,\beta +1;\gamma ;\varepsilon ) \\ 
-(\gamma -\beta )_{2}F_{1}(\alpha ,\beta -1;\gamma ;\varepsilon )=0,%
\end{array}
\label{32}
\end{equation}%
and letting $\alpha =\frac{k}{2}+1,$ $\beta =-n,$ $\gamma =q+1,$ $%
\varepsilon =2,$ we have%
\begin{eqnarray}
&&\left( q-k-1\right) \text{ }_{2}F_{1}(\frac{k}{2}+1,-n;q+1;2)+n\text{ }%
_{2}F_{1}(\frac{k}{2}+1,-n+1;q+1;2)  \label{33} \\
&&-(q+n+1)\text{ }_{2}F_{1}(\frac{k}{2}+1,-n-1;q+1;2)=0  \notag
\end{eqnarray}%
Substituting (30) into the right-hand side of (30) we finally obtain
(recovering $\mathfrak{C}(k))$ 
\begin{eqnarray}
&&\left( \xi \frac{\partial }{\partial \xi }+\xi ^{\ast }\frac{\partial }{%
\partial \xi ^{\ast }}+1\right) \left\langle \xi \right\vert \left.
q,k\right\rangle   \label{34} \\
&=&(q-k-1)\mathfrak{C}(k)e^{-|\xi |^{2}/2}\sum_{n=0}^{\infty }\frac{1}{n!}%
H_{n+q,n}\left( \xi ^{\ast },\xi \right) _{2}F_{1}(-n,\frac{k}{2}+1;q+1;2) 
\notag \\
&=&(q-k-1)\left\langle \xi \right\vert \left. q,k\right\rangle ,  \notag
\end{eqnarray}%
thus we have proved the solution of (14). On the other hand, from (21) we
know 
\begin{equation}
H_{n+q,n}\left( \xi ^{\ast },\xi \right) =e^{-iq\varphi }H_{n+q,n}\left(
|\xi |,|\xi |\right) ,  \label{35}
\end{equation}%
so the solution (14) automatically satisfied with (16). The solution seems
new.

\section{The simultaneous eigenstate of $Q$ and $\left( ab-a^{\dagger
}b^{\dagger }\right) $}

Now we hope to obtain $\left\vert q,k\right\rangle $ from the information of 
$\left\langle \xi \right\vert \left. q,k\right\rangle .$ Using the
completeness relation (9) of $\left\vert \xi \right\rangle $ and (17)-(18)
we can have 
\begin{eqnarray}
\left\vert q,k\right\rangle  &=&\int \frac{d^{2}\xi }{\pi }\left\vert \xi
\right\rangle \left\langle \xi \right\vert \left. q,k\right\rangle 
\label{36} \\
&=&\mathfrak{C}(k)\int \frac{d^{2}\xi }{\pi }\left\vert \xi \right\rangle
e^{-|\xi |^{2}/2}\sum\limits_{n=0}^{\infty }\frac{1}{n!}H_{n+q,n}\left( \xi
^{\ast },\xi \right) \text{ }_{2}F_{1}(-n,\frac{k}{2}+1;q+1;2).  \notag
\end{eqnarray}%
Then using (22) we expand $\left\vert \xi \right\rangle $ in (5),%
\begin{eqnarray}
\left\vert \xi \right\rangle  &=&e^{-|\xi |^{2}/2}\sum_{l,j=0}^{\infty }%
\frac{a^{\dagger l}b^{\dagger }{}^{j}}{l!j!}H_{l,j}(\xi ,\xi ^{\ast
})\left\vert 0,0\right\rangle   \label{37} \\
&=&e^{-|\xi |^{2}/2}\sum_{l,j=0}^{\infty }\frac{1}{\sqrt{l!j!}}H_{l,j}\left(
\xi ,\xi ^{\ast }\right) \left\vert l,j\right\rangle ,  \notag
\end{eqnarray}%
where $\left\vert l,j\right\rangle $ is the two-mode Fock state.
Substituting (37) into (36) and using the integration formula%
\begin{equation}
\int \frac{d^{2}\xi }{\pi }e^{-|\xi |^{2}}H_{l,j}\left( \xi ,\xi ^{\ast
}\right) H_{m,n}^{\ast }\left( \xi ,\xi ^{\ast }\right) =\delta _{l,m}\delta
_{j,n}m!n!,  \label{38}
\end{equation}%
we have 
\begin{eqnarray}
\left\vert q,k\right\rangle  &=&\mathfrak{C}(k)\sum\limits_{n=0}^{\infty }%
\frac{1}{n!}_{2}F_{1}(-n,\frac{k}{2}+1;q+1;2)  \label{39} \\
&&\times \int \frac{d^{2}\xi }{\pi }e^{-|\xi |^{2}}\sum_{l,j=0}^{\infty }%
\frac{1}{\sqrt{l!j!}}H_{l,j}\left( \xi ,\xi ^{\ast }\right) H_{n+q,n}^{\ast
}\left( \xi ,\xi ^{\ast }\right) \left\vert l,j\right\rangle   \notag \\
&=&\mathfrak{C}(k)\sum_{n=0}^{\infty }\frac{\sqrt{\left( n+q\right) !}}{%
\sqrt{n!}}\text{ }_{2}F_{1}(-n,\frac{k}{2}+1;q+1;2)\left\vert
n+q,n\right\rangle   \notag \\
&=&\mathfrak{C}(k)a^{\dagger q}\sum_{n=0}^{\infty }\frac{a^{\dagger
n}b^{\dagger n}}{n!}\text{ }_{2}F_{1}(-n,\frac{k}{2}+1;q+1;2)\left\vert
0,0\right\rangle .  \notag
\end{eqnarray}%
As we have known that the convergent condition for the hypergeometric
function defined in (19) is $|z|<1,$ $\gamma \neq 0,-1,-2,\cdot \cdot \cdot ,
$ so $\left\vert q,k\right\rangle $ seems not normalized as a finite number,
but as a divergent one. To see this more clearly, using (17)-(18), (9) and
(38), formally we have 
\begin{eqnarray}
\left\langle q,k\right\vert \left. q,k\right\rangle  &=&\int \frac{d^{2}\xi 
}{\pi }\left\langle q,k\right\vert \left. \xi \right\rangle \left\langle \xi
\right\vert \left. q,k\right\rangle   \label{40} \\
&=&|\mathfrak{C}(k)|^{2}\int \frac{d^{2}\xi }{\pi }\sum_{n=0}^{\infty }\frac{%
1}{n!}H_{n+q,n}\left( \xi ^{\ast },\xi \right) \text{ }_{2}F_{1}(-n,\frac{k}{%
2}+1;q+1;2)  \notag \\
\times  &&\sum_{n^{\prime }=0}^{\infty }\frac{1}{n^{\prime }!}H_{n^{\prime
}+q,n^{\prime }}\left( \xi ,\xi ^{\ast }\right) \text{ }\left[
_{2}F_{1}(-n^{\prime },\frac{k}{2}+1;q+1;2)\right] ^{\ast }  \notag \\
&=&|\mathfrak{C}(k)|^{2}\sum_{n=0}^{\infty }\frac{\left( n+q\right) !}{n!}%
\text{ }|_{2}F_{1}(-n,\frac{k}{2}+1;q+1;2)|^{2}.  \notag
\end{eqnarray}%
Therefore, the common eigenvector of $Q$ and $\left( ab-a^{\dagger
}b^{\dagger }\right) ,$ like the common eigenvector of $Q$ and $a^{\dagger
}b^{\dagger },$ is normalized as a singular function, so its applications
are greatly limited. However, the exploration for formal solution of $%
\left\vert q,k\right\rangle $ has its own sense in mathematical physics.

In summary, as a continuum work of Bhaumik et al [3] we have set a complex
differential equation in the entangled state representation for deriving the
new the common eigenstate $\left\vert q,k\right\rangle $ of parametric
interaction Hamiltonian and number-difference operator, the complex
differential equation has been solved in terms of hypergeometric function
and the two-variable Hermite polynomials and the solution seems new. Thus
this paper embodies a new use of hypergeometric functions and is of
theoretical mathematical physics meaning. This work is also an addendum to
Ref. [5], in which the common eigenkets of $Q$ and two-mode creators $%
a^{\dagger }b^{\dagger }$ are discussed.

\end{document}